\documentclass[aps,pra,twocolumn,superscriptaddress]{revtex4-1}
\usepackage{graphicx}
\usepackage{amssymb}
\usepackage{subfigure}
\usepackage{amsmath}
\usepackage{amsfonts}
\usepackage{physics}
\usepackage{bm}
\usepackage{color}
\usepackage{comment}
\usepackage[dvipsnames]{xcolor}
\usepackage[normalem]{ulem}

\newcommand{\vect}[1]{\ensuremath{\bm{\mathrm{#1}}}}

\newcommand{\oper}[1]{\mathbf{\mathsf{#1}}}

\newcommand{\brm}[1]{\ensuremath{\mathbf{#1}}}

\newcommand{\sinc}{\ensuremath{\mathrm{sinc}}}

\definecolor{armygreen}{rgb}{0.29, 0.33, 0.13}

\begin{document}
\title{Quantum-optical Description of Phase Conjugation of Vector Vortex Beams in Stimulated Parametric Down Conversion}
\author{A. G. de Oliveira}
\affiliation{Departamento de F\'{i}sica, Universidade Federal de Santa Catarina, CEP 88040-900, Florian\'{o}plis, SC, Brazil}
\author{N. Rubiano da Silva}
\affiliation{Departamento de F\'{i}sica, Universidade Federal de Santa Catarina, CEP 88040-900, Florian\'{o}plis, SC, Brazil}
\author{R. Medeiros de Ara\'{u}jo}
\affiliation{Departamento de F\'{i}sica, Universidade Federal de Santa Catarina, CEP 88040-900, Florian\'{o}plis, SC, Brazil}
\author{P. H. Souto Ribeiro}
\email{p.h.s.ribeiro@ufsc.br}
\affiliation{Departamento de F\'{i}sica, Universidade Federal de Santa Catarina, CEP 88040-900, Florian\'{o}plis, SC, Brazil}
\author{S. P. Walborn}
\affiliation{Instituto de F\'{\i}sica, Universidade Federal do Rio de Janeiro, Caixa Postal 68528, Rio de Janeiro, RJ 21941-972, Brazil}
\affiliation{Departamento de F\'{\i}sica, Universidad de Concepci\'on, 160-C Concepci\'on, Chile}
\affiliation{Millennium Institute for Research in Optics, Universidad de Concepci\'on, 160-C Concepci\'on, Chile}
\begin{abstract}
We present a quantum optics approach for describing stimulated parametric down-conversion in the two type-I-crystal ``sandwich''configuration, which allows for parametric interaction of vector vortex beams. We analyze the conditions for which phase conjugation of the seed vector beam occurs.  We then use two strategies for defining generalized Stokes parameters to describe phase conjugation of vector vortex beams. These allow for geometrical representations, such as higher-order Poincar\'e spheres.  Our results are useful for description and design of stimulated and spontaneous parametric down conversion experiments with vector vortex beams.
\end{abstract}
\maketitle
\section{Introduction}

Parametric interactions are nonlinear processes that describe energy exchange between oscillators \cite{Siegman66}. In the optical regime, they may describe energy exchange between optical waves of different frequencies \cite{Yariv65,Yariv66}. Among the optical parametric processes, very much attention has been paid to  parametric down conversion, particularly in the spontaneous regime \cite{burnham70,mandel95}, known as SPDC.  On the other hand, when one of the down-converted modes is seeded with a non-vacuum field such as a laser beam, stimulated parametric down conversion (StimPDC) occurs\cite{wang91}.  StimPDC can be used to explore interesting processes such as phase conjugation \cite{ribeiro01b}, or produce nonclassical states of light \cite{zavatta04,Kolkiran08,barbieri10,kiesel11}. 
\par
In SPDC, the quantum state of the down-converted photons can be engineered using different combinations of sources.  A widely employed example is the two-type-I crystal "sandwich" source, which was conceived to prepare post-selection-free two-photon polarization states with a variable degree of entanglement \cite{kwiat99}.  StimPDC in this two type-I-crystal configuration has been used to efficiently probe the polarization entanglement of the source in the spontaneous regime \cite{rozema15}. 
\par
  More recently, it has been shown that the two-crystal sandwich source allows for StimPDC with vector vortex beams (VVBs) \cite{Rosales18}, which can be used both as the pump or the seed (stimulating) beam \cite{Oliveira20}. This is possible because the two crystals mediate the coherent interaction for two orthogonal polarization directions. Since these two orthogonal polarizations form a basis for the polarization state of paraxial beams, each polarization component of the pump and seed beams can participate in the down-conversion process.  
StimPDC with VVBs is an interesting subject due to the possibility of combining the broad range of applications of VVBs in fields such as optical communications \cite{souza08,dambrosio12,Farias15,Zhao15,Milione15,Milione15A,Ndagano15,zhang16,Li16,Ndagano18}, metrology \cite{dambrosio13b}, imaging \cite{Biss06,zhan09,Yoshida19} and optical simulation of quantum systems \cite{oliveira05,souza07,borges10,passos18}, together with features such as phase conjugation \cite{Oliveira19,Oliveira20} and frequency conversion.
\newline
\hspace*{0.25cm}
The purpose of this paper is to present a theoretical approach based on a quantum-optics formalism that allows for the description of two-type-I-crystal StimPDC with VVBs, which to our knowledge has been lacking until now. Previous descriptions of StimPDC of transverse spatial modes  \cite{soutoribeiro99,caetano02,Oliveira19} do not take into account the polarization degree of freedom (DoF), where much of the interesting and particular features lie.  In section \ref{sec:III} we first discuss StimPDC of plane waves with this source, focusing only on the polarization degree of freedom.  We show how the entangled state that the source would produce in SPDC determines the polarization transformation from seed to idler beam in StimPDC.  In section \ref{sec:IV} we provide a theoretical quantum-optics-based description of VVBs and then StimPDC with VVBs.  We determine both the density operator and the detection probability (intensity) associated with the stimulated idler beam, showing that these correspond to the phase conjugation of the input seed VVB.  In sections \ref{sec:V} and \ref{sec:VI} we describe phase conjugation of VVBs using generalized Stokes vector formalisms.  The first focuses on cylindrically symmetric vector beams, and makes use of a geometrical description based on the higher-order Stokes parameters and the higher-order Poincar\'e Sphere \cite{milione2011higher,milione2012higher} to visualize the phase conjugation of VVBs. The second approach employs a general formalism analogous to two qubits that is applicable to any vector beam.   Conclusions and avenues for future work are discussed in section \ref{sec:VII}.

\section{Preliminary concepts}
\label{sec:II}
\subsection{Phase conjugation of polarization states and Stokes parameters}
Let us consider a generic polarization pure state written with respect to the basis formed by $\ket{H}$ and $\ket{V}$, representing horizontal and vertical polarization states, respectively:
\begin{equation}
\ket{\theta,\phi} = \cos \frac{\theta}{2} \ket{H} + e^{i \phi} \sin\frac{\theta}{2} \ket{V}. 
\label{eq:pure-state}
\end{equation}
Mathematically, the conjugation of the above polarization state is defined as its complex conjugate: 
\begin{equation}
\ket{\theta,\phi}^* = \cos \frac{\theta}{2} \ket{H} + e^{-i \phi} \sin\frac{\theta}{2} \ket{V} = \ket{\theta,-\phi}.
\end{equation}
Note that, following this definition, the states $\ket{H}$, $\ket{V}$, $\ket{D}$ and $\ket{A}$ ($D$ and $A$ standing for diagonal and antidiagonal polarizations, respectively) are their own conjugates, while the circularly polarized states $\ket{R}=(\ket{H}+i\ket{V})/\sqrt{2}$ and $\ket{L}=(\ket{H}-i\ket{V})/\sqrt{2}$ are conjugates of one another.

Conjugation of the polarization state of light may also be described in terms of the Stokes vector $\vec{\mathbf{S}}=(S_0,S_1,S_2,S_3)^T$. The Stokes parameters $S_i$ are defined in the following way: $S_0=I$ is the total intensity of the beam; $S_1 =I_H-I_V$, $S_2=I_D-I_A$ and $S_3=I_R-I_L$, where $I_\varepsilon$ stands for the remaining intensity after projection onto the state $\ket{\varepsilon}$. We may assume the beam intensity is normalized to 1 ($S_0=1$), so that the vector $\vec{S}=(S_1,S_2,S_3)^T$ is sufficient to characterize the polarization state. Under this condition, the degree of polarization is the euclidean norm of $\vec{S}$: $p=\sqrt{S_1^2+S_2^2+S_3^2}$. For pure states, such as Eq.\,(\ref{eq:pure-state}), the degree of polarization is maximum from definition, i.e. $p=1$.

Henceforth, we will refer to $\vec{S}$ as the Stokes vector, for simplicity. For state $\ket{\theta,\phi}$, the Stokes vector reads
\begin{equation}
\vec{S}=(\cos\theta,\ \sin \theta \cos\phi,\ \sin \theta \sin \phi)^T,
\label{eq:stokesvec}
\end{equation}   
while, for the conjugate state $\ket{\theta,\phi}^*$, we have 
\begin{equation}
\vec{S}^*=(\cos\theta,\ \sin \theta \cos\phi,\ -\sin \theta \sin \phi)^T. 
\label{stokes}
\end{equation} 
From Eq.\,(\ref{stokes}), one can see that conjugation simply changes the sign of the $S_3$ (or $R/L$) component of the Stokes vector: $\vec{S}^*=(S_1,S_2,-S_3)^T$. Therefore, in terms of the Poincar\'e sphere of polarization, which has the $\ket{R}$ ($\ket{L}$) state sitting on its North (South) pole, conjugation corresponds to a mirror reflection through the equatorial plane of the sphere \cite{Oliveira20}. 

\par
\section{Vector Phase Conjugation in StimPDC and Bell states}
\label{sec:III}

\begin{figure}
    \centering
    \includegraphics[width=\columnwidth]{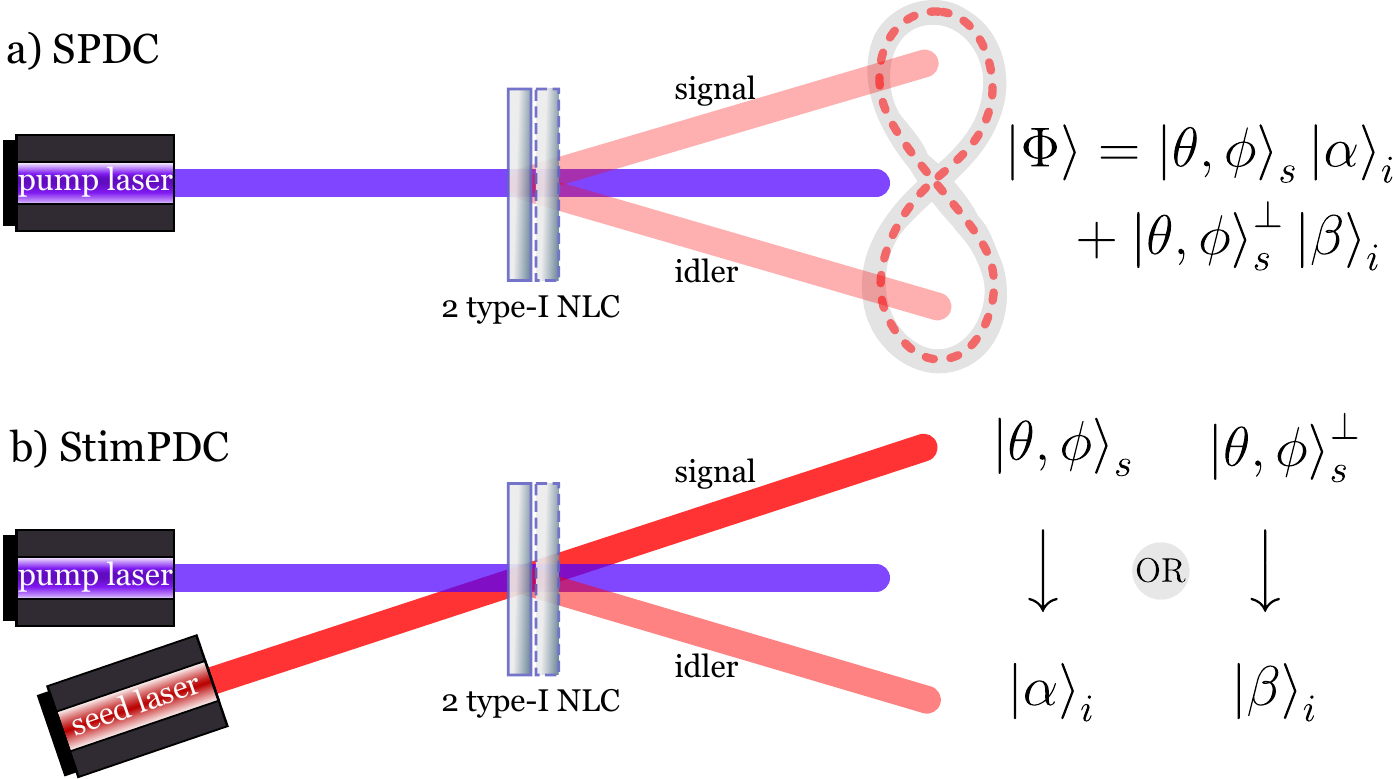} 
     \caption{(Color online) a) SPDC with two type-I crystals generating polarization entangled states. b) StimPDC with two type-I crystals: the seed laser polarization determines the idler polarization. NLC: nonlinear crystal.} 
       \label{fig:stimpdcsource}
\end{figure}
The Bell basis of entangled polarization states is composed of four states. An SPDC source can be set in four different manners that creates each of the Bell states, or even produce superpositions of these states. Let us see how the vector phase conjugation appears in those scenarios.

Consider the two-crystal type-I SPDC source, as shown in Fig.\,\ref{fig:stimpdcsource} a), with an intense pump beam that has the polarization state $\ket{\theta_p,\phi_p}$.  Assume also that all fields are monochromatic and single spatial mode, which can be achieved via pump engineering and filters.  In this simple scenario, the effective SPDC Hamiltionian is  $\propto \cos (\theta_p/2) \oper{a}_V^\dagger\oper{a}_V^\dagger + \exp({i \phi_p}) \sin(\theta_p/2) \oper{a}_H^\dagger\oper{a}_H^\dagger$, which creates two photons with the same polarization. 
For SPDC, where the signal and idler modes are initially in the vacuum state, it will produce the state
\begin{equation}
\ket{\Phi}= \sin\frac{\theta_p}{2} e^{i\phi_p }\ket{H}_s\ket{H}_i + \cos\frac{\theta_p}{2}\ket{V}_s\ket{V}_i, 
\label{eq:SPDCstate}
\end{equation}
where it is clear that the pump beam polarization determines the entangled state. One can also consider a source that produce states of the form
\begin{equation}
\ket{\Psi}= \sin\frac{\theta_p}{2} e^{i\phi_p }\ket{H}_s\ket{V}_i + \cos\frac{\theta_p}{2}\ket{V}_s\ket{H}_i = \oper{X}_i\ket{\Phi},
\label{eq:SPDCstate2}
\end{equation}
where $\oper{X}_i$ is the Pauli operator (half-wave plate at $45^\circ$), that changes $H \leftrightarrow V$. The same two-crystal type-I source with an additional half-wave plate can produce this state, or alternatively, type-II crossed cones \cite{kwiat95} and Sagnac source \cite{Kim06}, for instance, 

Now, let us rewrite the states of Eqs.\,\eqref{eq:SPDCstate} and \eqref{eq:SPDCstate2} in a more general way using the following \textit{orthonormal basis} for the signal photons: $\{\ket{\theta,\phi}_s, \ket{\theta,\phi}_s^\perp\}$, where $\ket{\theta,\phi}_s$ is defined as in Eq.\,\eqref{eq:pure-state} and 
\begin{equation}
\ket{\theta,\phi}_s^\perp \equiv \ket{\theta-\pi,\phi}_s = \sin \frac{\theta_s}{2} \ket{H}_s - e^{i \phi_s} \cos \frac{\theta_s}{2} \ket{V}_s.  
\end{equation}
The results are
\begin{equation}
\ket{\Phi}=\ket{\theta,\phi}_s\ket{\alpha}_i + \ket{\theta,\phi}^\perp_s\ket{\beta}_i 
\label{eq:SPDCstate2phi}
\end{equation}
and 
\begin{equation}
\ket{\Psi}=\ket{\theta,\phi}_s\oper{X}_i\ket{\alpha}_i + \ket{\theta,\phi}^\perp_s\oper{X}_i\ket{\beta}_i, 
\label{eq:SPDCstate2psi}
\end{equation}
where
\begin{align}
\ket{\alpha}_i&= \sin\frac{\theta_p}{2} \cos\frac{\theta_s}{2} \ket{H}_i + e^{-i(\phi_s-\phi_p)} \cos\frac{\theta_p}{2} \sin\frac{\theta_s}{2} \ket{V}_i\nonumber\\
\ket{\beta}_i&= \sin\frac{\theta_p}{2} \sin\frac{\theta_s}{2} \ket{H}_i - e^{-i(\phi_s-\phi_p)} \cos\frac{\theta_p}{2} \cos\frac{\theta_s}{2} \ket{V}_i.
\end{align}
\par
Suppose now that we have StimPDC with the  seed beam aligned to the signal arm, matched in path and frequency to the photon $s$. Moreover, we assume that the intensity of the seed beam is high enough, so that the intensity of the idler beam is dominated by the stimulated process and the spontaneous emission contribution is negligible. Fig.\,\ref{fig:stimpdcsource} b) illustrates the StimPDC scheme.  It has been shown that the stimulation of the signal beam can be interpreted as being equivalent to a state projection \cite{Arruda18}. Therefore, we can see from Eqs.\,\eqref{eq:SPDCstate2phi} and \eqref{eq:SPDCstate2psi} that, if the seed beam is prepared in the arbitrary polarization state $\ket{\theta,\phi}_s$, then the stimulated beam in the idler arm will leave the crystals in polarization state $\ket{\alpha}_i$ or $\oper{X}_i\ket{\alpha}_i$, depending on which type of source is used. These results may be summarized as $\oper{X}_i^{t}\ket{\alpha}_i$, where $t=0$ (1) when the source produces a state of type $\Phi$ ($\Psi$). 

The effect of the Pauli operator $\oper{X}$ on an arbitrary Stokes vector $\vec{S}$ is given by ${X}[(S_1,S_2,S_3)^T] = (-S_1,S_2,-S_3)^T$, since it interchanges both $H\leftrightarrow V$ and $R\leftrightarrow L$. In Ref. \cite{Oliveira20} the Stokes vector were derived for StimPDC with source $\Phi$ and a seed beam described by polarization state $\ket{\alpha}$.  We can generalize these results to include the $\Psi$-type source.  Namely, the Stokes vector of the stimulated beam can be written as
\begin{align}
    \Vec{S}_{i,t} =\frac{1}{2}\left( 
    \begin{matrix}
       (-1)^t( S_{s1}-S_{p1}) \\
        S_{p2}S_{s2}-S_{p3}S_{s3} \\
   - (-1)^t (S_{s2}S_{p3}+S_{p2}S_{s3})
    \end{matrix}
    \right),
    \label{eq:idler-stokes}
\end{align}
where $S_{pj}$ are the Stokes parameters for the pump beam, with $j = 1,2,3$.
\par
It is interesting to see what happens when the source is set to produce the four Bell states $\ket{\Phi^{\pm}}$ and $\ket{\Psi^{\pm}}$.   For StimPDC, the idler polarization for these four cases is: 
\begin{equation}
\vec{S}_i^{\Phi^+}=(S_{s1},S_{s2},-S_{s3}),
\label{S3}
\end{equation}
which is the usual vector phase conjugation; 
\begin{equation}
\vec{S}_i^{\Phi^-}=(S_{s1},-S_{s2},S_{s3}),
\label{S2}
\end{equation}
which is ``phase conjugation" but with respect to the $S_2$ axis; \cite{Oliveira20}, 
\begin{equation}
\vec{S}_i^{\Psi^+}=(-S_{s1},S_{s2},S_{s3}),
\label{S1}
\end{equation}
which is ``phase conjugation" but with respect to the $S_1$ axis; and
\begin{equation}
\vec{S}_i^{\Psi^-}=(-S_{s1},-S_{s2},-S_{s3}),
\end{equation}
which is a universal NOT operation, inverting the polarization state, as has been observed at the single photon level \cite{demartini02,demartini04}, where the fidelity of the NOT operation is reduced by the significant contribution of the spontaneous emission. 

Recalling that phase conjugation depends on the basis chosen to represent the state, we interpret the results in Eqs. (\ref{S3}), (\ref{S2}) and (\ref{S1}) as phase conjugation in different bases.
 
\section{StimPDC and phase conjugation of vector vortex beams}
\label{sec:IV}

\begin{figure}
    \centering
    \includegraphics[width=\columnwidth]{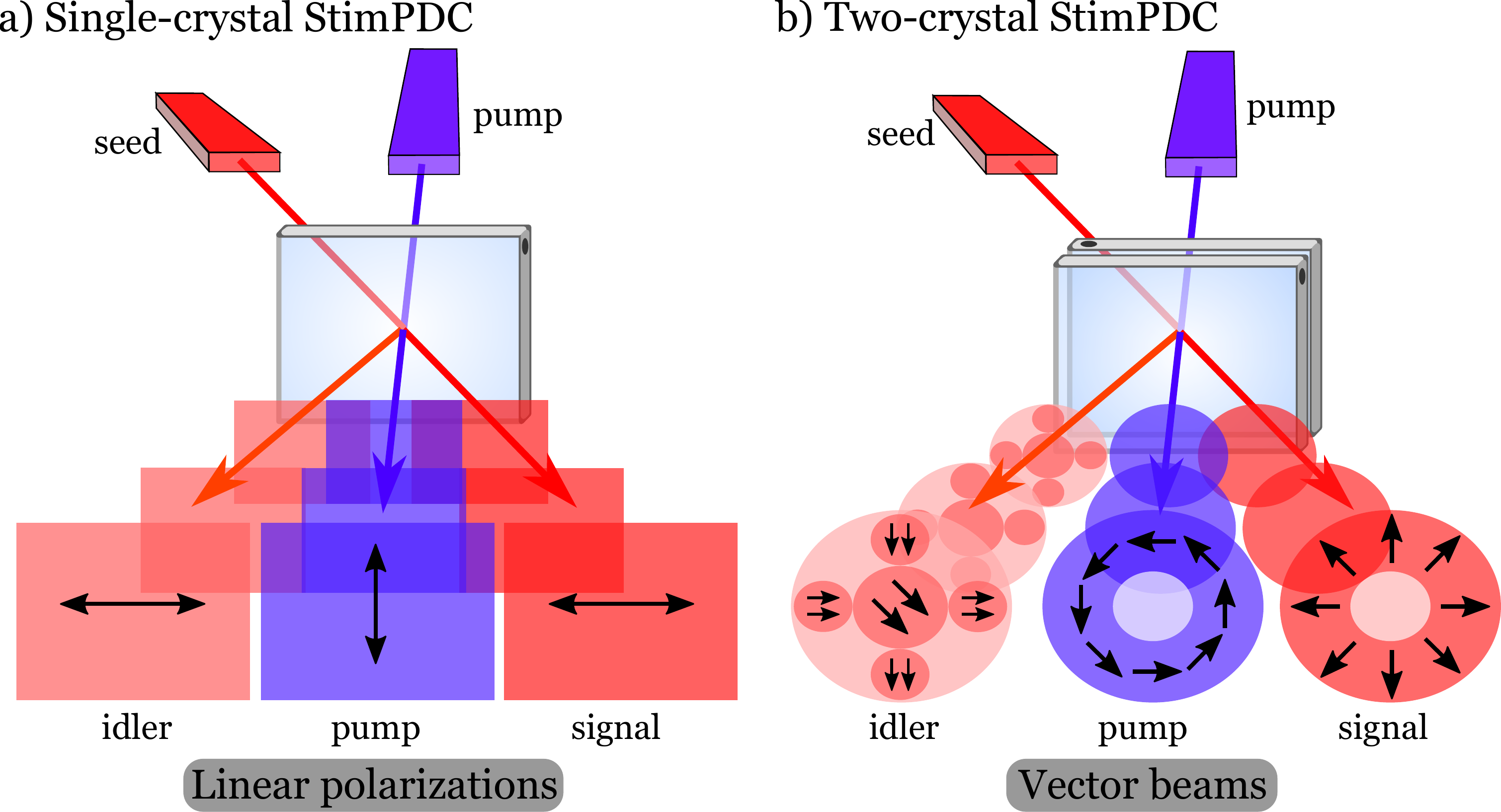} 
    \caption{(Color online) Sketch of the StimPDC scheme for a) single-crystal source and plane waves, b) two-type-I-crystal source and vector beams.} 
    
       \label{fig2}
\end{figure}

In this section we will construct a more complete and more general theoretical description of StimPDC, which will allow for the description of StimPDC with VVB.  First, it will be useful to establish the quantum-optical description of coherent states and single-photon states in optical modes describing vector beams. 
The experimental scheme considered here is sketched in Fig.\,\ref{fig2}, which shows in Fig.\,\ref{fig2} a) the usual scheme with one crystal and plane waves (for simplicity) and in Fig.\,\ref{fig2}b) the two-type-I-crystal scheme with VVB.

\subsection{Quantum-optical description of vector beams}
In classical optics, a paraxial, monochromatic vector beam can be written as  
\begin{equation}
\psi_{vb}(\brm{q})=\gamma_H  \vect{e}_H \psi_H (\brm{q})+ \gamma_V   \vect{e}_V \psi_V (\brm{q}),
\label{eq:vb}
\end{equation}
where $ \vect{e}_j$ are unit polarization vectors in the $j=H,V$ directions, $\gamma_j$ are complex numbers obeying $|\gamma_H|^2+|\gamma_V|^2=1$, and $\psi_j$ describe orthogonal spatial modes in terms of transverse momentum coordinates $\brm{q}$.  
\par
Through quantization of Eq.\,\eqref{eq:vb}, we can define the operator
\begin{equation}
\oper{a}^\dagger_{vb} = \gamma_H \oper{a}^\dagger_{H,\psi_H} +\gamma_V \oper{a}^\dagger_{V,\psi_V},
\label{eq:avb}
\end{equation}
where $\oper{a}^\dagger_{j,\psi_j}$ creates a photon with polarization $j$ and transverse spatial mode $\psi_j$. 

A coherent state with mean photon number $|\alpha|^2$, prepared in a vector beam mode, is denoted by $\ket{\alpha}_{vb}$.  
This can be written in terms of creation operators as 
\begin{equation}
\ket{\alpha}_{vb} = e^{-\frac{|\alpha|^2}{2}} e^{\alpha \oper{a}^\dagger_{vb}}\ket{vac}. 
\label{eq:alphavb}
\end{equation}

Using operator of Eq.\,\eqref{eq:avb}, the state of Eq.\,\eqref{eq:alphavb} can be rewritten 
\begin{align}
\ket{\alpha}_{vb} & = e^{-\frac{|\alpha|^2}{2}} e^{\gamma_H \alpha \oper{a}^\dagger_{H,\psi_H}}e^{\gamma_V \alpha \oper{a}^\dagger_{V,\psi_V}}\ket{vac} \nonumber \\
& = \ket{\gamma_H \alpha}_{H,\psi_H}\ket{\gamma_V \alpha}_{V,\psi_V}, 
\label{eq:alphavb2}
\end{align}
where in the second line we used  $|\gamma_H|^2+|\gamma_V|^2=1$. 
\par
In a similar fashion, we can expand the transverse mode operators in terms of plane wave modes,   
\begin{equation}
 \oper{a}^\dagger_{j,\psi_j} = \sum_{\brm{q}}\psi_j(\brm{q})\oper{a}_{j,\brm{q}}.
 \label{eq:apw}
 \end{equation}
 In this way, a coherent state of the form $\ket{\eta}_{j,\psi_j}$ is
\begin{align}
\ket{\eta}_{j,\psi_j} & = e^{-\frac{|\eta|^2}{2}} e^{\eta \oper{a}^\dagger_{j,\psi_j}}\ket{vac} \nonumber \\
& = \prod_{\brm{q}}e^{-\frac{|\eta|^2}{2}} e^{\eta \psi_j(\brm{q}) \oper{a}^\dagger_{j,\brm{q}}}\ket{vac} \nonumber \\
& = \prod_{\brm{q}} \ket{\eta \psi_j(\brm{q})}_{j,\brm{q}}. 
\label{eq:alphaj}
\end{align}
Finally, using Eq.\,\eqref{eq:alphaj}, we can write the vector beam of Eq.\,\eqref{eq:alphavb2} in terms of plane wave modes as 
\begin{equation}
\ket{\alpha}_{vb} = \prod_{\brm{q}} \ket{ \alpha \gamma_H  \psi_H(\brm{q})}_{H,\brm{q}} \ket{\alpha \gamma_V  \psi_V(\brm{q})}_{V,\brm{q}}. 
\label{eq:alphavb3}
\end{equation}
Eq.\,(\ref{eq:alphavb3}) will be necessary for the next section. 
\par
Using operator of Eq.\,\eqref{eq:avb}, a single photon in a vector beam mode is  
\begin{equation}
\ket{1}_{vb}=\oper{a}^\dagger_{vb} \ket{vac} = \gamma_H \ket{1}_{H,\psi_H} +\gamma_V \ket{1}_{V,\psi_V}, 
\label{eq:1vb}
\end{equation}
which can be recast in terms of plane-wave modes using Eq.\,\eqref{eq:apw}:  
\begin{equation}
\ket{1}_{vb}=  \gamma_H  \int d{\brm{q}} \psi_H(\brm{q})  \ket{1}_{H,\brm{q}} +\gamma_V  \int d{\brm{q}} \psi_V(\brm{q}) \ket{1}_{V,\brm{q}}, 
\label{eq:1vb2}
\end{equation}
where we extend the summations to integrals. 
We can note the key difference between the vector beam composed of a single photon in Eq.\,\eqref{eq:1vb2} and the coherent state of Eq.\,\eqref{eq:alphavb3} written in the plane-wave basis. While the latter is clearly a product state of all relevant modes, the single photon state of Eq.\,\eqref{eq:1vb2} is an example of mode entanglement. 
\par
For comparison with the next section, let us note that the phase conjugate of this state is 
\begin{equation}
\ket{1}^{*}_{vb}=  \gamma_H^*  \int d{\brm{q}} \psi^*_H(\brm{q})  \ket{1}_{H,\brm{q}} +\gamma^*_V  \int d{\brm{q}} \psi^*_V(\brm{q}) \ket{1}_{V,\brm{q}}. 
\label{eq:1vbpc}
\end{equation}

\subsection{StimPDC of Vector Vortex Beams}
Let us now turn to StimPDC of vector beams using a PDC source consisting of two type-I crystals as in Fig.\,\ref{fig:stimpdcsource} b).  We assume that the monochromatic and paraxial approximations are valid for all fields, and that higher order photon number terms (four or more SPDC photons) are negligible. We will first consider a single down-conversion crystal that converts energy from a $H$ polarized pump field into down-converted fields with $V$ polarizations.  As is usual, we assume that transverse width of the crystal is much larger than that of the transverse width of the pump beam. For our purposes, it is sufficient to consider the non-linear interaction term described by \cite{walborn10,schneeloch16} 
\begin{equation}
\iiint d\brm{q}_p\, d\brm{q}_s\, d\brm{q}_i F(\brm{q}_s, \brm{q}_i,\brm{q}_p)  \oper{a}_{V,\brm{q}_i}^\dagger \oper{a}_{V,\brm{q}_s}^\dagger\oper{a}_{H,\brm{q}_p},
\end{equation}
where $F(\brm{q}_s, \brm{q}_i,\brm{q}_p)= \delta(\brm{q}_s+ \brm{q}_i-\brm{q}_p) \sinc[(k_i\brm{q}_s/k_s-k_s\brm{q}_i/k_i)^2d/4{k_p}]$ is the phase matching function, $d$ is the length of the non-linear crystal, and $k_s,k_i, k_p$ are the wave numbers of the signal, idler and pump fields, respectively.  Here we assume that the coupling between the polarization and spatial/spectral degrees of freedom can be ignored.  This requires a thin crystal, or additional phase compensation using birrefringent elements \cite{rangarajan09}. 
Applying this operator to an intense pump beam that is described by a VVB $\ket{\alpha_p}_{vb}$ as in Eq.\,\eqref{eq:alphavb3}, we can write an effective Hamiltonian that acts only on the down-converted modes: 
\begin{equation}
\oper{H}_H   \approx  \gamma_{H} \iint  d\brm{q}_s d\brm{q}_i  \Psi_H(\brm{q}_s,\brm{q}_i) \oper{a}_{V,\brm{q}_s}^\dagger\oper{a}_{V,\brm{q}_i}^\dagger, 
\label{eq:HamH}
\end{equation}
where 
\begin{equation}
\Psi_H(\brm{q}_s,\brm{q}_i) = \psi_{H}(\brm{q}_s+\brm{q}_i) \sinc\left [\frac{d}{4{k_p}}\left (\frac{k_i}{k_s} \brm{q}_s-\frac{k_s}{k_i}\brm{q}_i\right )^2\right ],
\label{eq:PsiH}
\end{equation} 
and $\psi_{H}$ describes the angular spectrum of the pump beam.

In the two crystal source, a second crystal converting the $V$ polarization component of the  pump field to $H$-polarized down-converted fields is used, for which expressions analogous to Eqs.\,\eqref{eq:HamH} and \eqref{eq:PsiH} can be written (with $H \leftrightarrow V$).  Moreover, let us assume that these crystals are thin, with $d$ much smaller than the Rayleigh range of the pump beam,  and placed very close together, as shown in Fig.\,\ref{fig:stimpdcsource} b).  Under these conditions, the propagation effects between crystals can be ignored.  With the above approximations, the effective Hamiltonian for parametric down-conversion with the two-crystal source and a VVB pump beam is 

\begin{align}
 \oper{H} \approx  & \,\gamma_{H} \iint  d\brm{q}_s d\brm{q}_i  \Psi_H(\brm{q}_s,\brm{q}_i) \oper{a}_{V,\brm{q}_s}^\dagger\oper{a}_{V,\brm{q}_i}^\dagger \nonumber \\
 & +  \gamma_{V} \iint  d\brm{q}_s d\brm{q}_i  \Psi_V(\brm{q}_s,\brm{q}_i) \oper{a}_{H,\brm{q}_s}^\dagger\oper{a}_{H,\brm{q}_i}^\dagger.
 \label{eq:Ham}
 \end{align}
 Here we remember that $\gamma_H$ and $\gamma_V$ are complex coefficients describing the horizontal and vertical polarization components of the pump vector beam, respectively. 
 In StimPDC \cite{soutoribeiro99} with a vector beam as the seed signal beam, the PDC Hamiltonian of Eq.\,\eqref{eq:Ham} is applied to the initial state  $ \ket{\alpha_s}_{sv} \ket{vac}$, where
 \begin{equation}
 \ket{\alpha_s}_{sv} =  \prod_{\brm{q}_s} \ket{ \alpha_s \delta_H  \phi_H(\brm{q}_s)}_{H,\brm{q}_s} \ket{\alpha_s \delta_V  \phi_V(\brm{q}_s)}_{V,\brm{q}_s},
 \label{eq:alphasv}
  \end{equation}
 is the quantum-optical description of the VVB of the form of Eq.\,\eqref{eq:vb}, where $\delta_H$ and $\delta_V$ are the complex coefficients of each polarization component, $|\delta_H|^2+|\delta_V|^2=1$ and the index $_{sv}$ stands for seed VVB.
 
 With this we can follow the usual steps and calculate the initial state using time-dependent perturbation theory \cite{walborn10,schneeloch16}, resulting in
 \begin{align}
\label{vector99}
&|\Psi\rangle \approx \ket{\alpha_p}_{vb} |vac\rangle + \\ \nonumber  & C \gamma_H \iint  d\brm{q}_s d\brm{q}_i  \Psi_H(\brm{q}_s,\brm{q}_i) \ket{1}_{V,\brm{q}_i}  \oper{a}_{V,\brm{q}_s}^\dagger \ket{\alpha_s}_{sv} + \\ \nonumber
 &  C \gamma_{V} \iint  d\brm{q}_s d\brm{q}_i  \Psi_V(\brm{q}_s,\brm{q}_i) \ket{1}_{H,\brm{q}_i} \oper{a}_{H,\brm{q}_s}^\dagger \ket{\alpha_s}_{sv},
\label{eq:state1}
\end{align}
where $C\ll1$ is a constant. 

 From now on we will assume that this state will be used to determine photon-counting probabilities or intensities, so that we can ignore the vacuum term and focus on the non-vacuum contribution.  In each term of Eq.\,(\ref{vector99}), the signal field is a multimode single-photon-added coherent state which, under proper conditions, can exhibit non-classical \cite{agarwal91, zavatta04} and non-Gaussian \cite{barbieri10} behavior.  
In general, the idler field in state of Eq.\,\eqref{vector99} can be entangled to the signal field \cite{serna16}.  If $|\gamma_H|=|\gamma_V|$, this entanglement can be maximal in the spontaneous regime ($|\alpha_s|^2=0$) \cite{kwiat99}, and vanishes only when the intensity of the seed beam is very large ($|\alpha_s|^2\gg1$).  To proceed without loss of generality, we focus on the properties of the stimulated idler field by calculating its reduced density operator by tracing over the polarization and spatial state of the signal beam. This gives
 \begin{align}
\hat{\varrho}_i   \approx   \iiiint  & d\brm{q}_s d\brm{q}_i d\brm{q}^\prime_s d\brm{q}^\prime_i \sum_{j,k=H,V} \gamma_{\bar{j}} \gamma^*_{\bar{k }}  \ket{j,\brm{q}_i} \bra{{k,\brm{q}^\prime_i} }  \times \\ \nonumber 
& {\Psi}_{\bar{j}}(\brm{q}_s,\brm{q}_i) {\Psi}^*_{\bar{k}}(\brm{q}^\prime_s,\brm{q}^\prime_i)  
 _{sv}\bra{\alpha_s}\oper{a}_{k,\brm{q}^\prime_s} \oper{a}^\dagger_{j,\brm{q}_s} \ket{\alpha_s}_{sv}
\label{eq:rhoi}
\end{align}
where $\bar{j}$ is the orthogonal polarization to $j$, and in an effort towards simplification we introduce the notation $\ket{j,\brm{q}_i}=\ket{1}_{j,\brm{q}_i}$ for single photon states.
 
Applying the commutation relation to the term
\begin{equation}
_{sv}\bra{\alpha_s}\oper{a}_{k,\brm{q}^\prime_s} \oper{a}^\dagger_{j,\brm{q}_s} \ket{\alpha_s}_{sv}
 = \delta_{\brm{k,q_s},\brm{j,q'_s}} + _{sv}\bra{\alpha_s} \oper{a}^\dagger_{j,\brm{q}_s} \oper{a}_{k,\brm{q}^\prime_s} \ket{\alpha_s}_{sv}
 \label{eq:comm}
\end{equation}
we obtain two terms, where the first one is related to the SPDC component and  
using Eq.\,\eqref{eq:alphasv}, the second term results in 
\begin{equation}
 _{sv}\bra{\alpha_s} \oper{a}^\dagger_{j,\brm{q}_s} \oper{a}_{k,\brm{q}^\prime_s} \ket{\alpha_s}_{sv} =  |\alpha_s|^2 \delta^*_{j}\delta_{k} 
 \phi^*_j(\brm{q}_s) \phi_k(\brm{q}^\prime_s).
\end{equation}
The (unnormalized) density operator of the polarization and spatial degrees of freedom (DOF) of the idler field is 
\begin{equation}
\hat{\varrho}_i = \hat{\varrho}_i^{spdc}+  |\alpha_s|^2 \hat{\varrho}_i^{stim},
\label{eq:rhogen}
\end{equation}
where the component arising from SPDC is 
\begin{align}
\hat{\varrho}_i^{spdc}   \approx   \iiiint  & d\brm{q}_s d\brm{q}_i d\brm{q}^\prime_s d\brm{q}^\prime_i \sum_{j,k=H,V} \gamma_{\bar{j}} \gamma^*_{\bar{k }}  \ket{j,\brm{q}_i} \bra{{k,\brm{q}^\prime_i} }  \times \\ \nonumber 
& {\Psi}_{\bar{j}}(\brm{q}_s,\brm{q}_i) {\Psi}^*_{\bar{k}}(\brm{q}^\prime_s,\brm{q}^\prime_i),  
 \label{eq:rhospdc}
\end{align} 
and the component arising from StimPDC is
 \begin{align}
\hat{\varrho}_i^{stim}   \approx   \iiiint  & d\brm{q}_s d\brm{q}_i d\brm{q}^\prime_s d\brm{q}^\prime_i \sum_{j,k=H,V} \gamma_{\bar{j}} \gamma^*_{\bar{k }}  \ket{j,\brm{q}_i} \bra{{k,\brm{q}^\prime_i} }  \times \\ \nonumber 
&  \delta^*_{j}\delta_{k}  {\Psi}_{\bar{j}}(\brm{q}_s,\brm{q}_i) {\Psi}^*_{\bar{k}}(\brm{q}^\prime_s,\brm{q}^\prime_i)  
 \phi^*_j(\brm{q}_s) \phi_k(\brm{q}^\prime_s).
\label{eq:rhoi 2}
\end{align}
 \par
 To write the density operator in a more compact form, let us define the single photon states
\begin{equation}
\ket{j,\Phi^*_j} = \iint d\brm{q}_s d\brm{q}_i {\Psi}_{\bar{j}}(\brm{q}_s,\brm{q}_i) \phi^*_j(\brm{q}_s) \ket{j,\brm{q}_i}, 
\label{eq:stimmodes}
\end{equation}
so that we can write the StimPDC component in terms of these states as  
\begin{equation}
\hat{\varrho}_i^{stim}  = \sum_{j,k=H,V} 
 \gamma_{\bar{j}} \gamma_{\bar{k}}^* \delta^*_j \delta_k \ket{j,\Phi^*_j} \bra{k,\Phi^*_k}. 
\label{eq:rhoi 3}
\end{equation}
We note again that $\hat{\varrho}_i^{stim}$ is not normalized. The idler states defined in Eq.\,\eqref{eq:stimmodes} and used to write the StimPDC contribution are determined by the transverse mode functions of the seed VVB, as well as the pump beam through the two-photon amplitudes as defined in Eq.\,\eqref{eq:PsiH}.  This will be discussed in more detail below. 

\subsection{Intensity distribution of idler field}
 Before discussing phase conjugation, let us briefly relate the above results to what is usually measured in the laboratory. 
In typical experiments, the VVB is observed by first performing a polarization projection, and then measuring the spatial distributions of each polarization state with a detector or camera. Let us consider this scenario, with projection onto state $\ket{\theta,\phi}$ defined in Eq.\,\eqref{eq:pure-state}.   We note that the total intensity can be obtained by simply summing the intensities for two orthogonal polarization projections. To determine the detection probability or intensity of the idler field, we can then calculate  
\begin{equation}
    I_{\theta,\phi}(\brm{r}_i) \propto \mathrm{tr} \left[\oper{E}^-(\brm{r}_i,t)\oper{E}^+(\brm{r}_i,t) \bra{\theta,\phi}\hat{\varrho}_i\ket{\theta,\phi} \right], 
    \label{eq:I}
    \end{equation}
 which can be rewritten in terms of spontaneous and stimulated contributions using Eq.\,\eqref{eq:rhogen}:
 \begin{equation}
    I_{\theta,\phi}(\brm{r}_i) = I_{\theta,\phi}^{spdc}(\brm{r}_i) + |\alpha_s|^2 I_{\theta,\phi}^{stim}(\brm{r}_i).  
    \label{eq:I2}
    \end{equation}
 For paraxial propagation, the field operators corresponding to detection at position $\brm{r}=(x,y,z)$ (assuming free propagation from $z=0$ to $z$) can be written as
 \begin{equation}
     \oper{E}^+(\brm{r}) \propto \int d\brm{q}  e^{-i \brm{q}\cdot\brm{r}}e^{i q^2z/2k} \oper{a}_{\brm{q}},  
 \end{equation}
    and we can calculate the StimPDC component
     \begin{align}
I_{\theta,\phi}^{stim}(\brm{r}_i) = &
\cos^2\frac{\theta}{2} |\gamma_{V}|^2 |\delta_H|^2 |\oper{E}^+(\brm{r}) \ket{\Phi^*_H}|^2   + \nonumber  \\
 &  \sin^2\frac{\theta}{2} |\gamma_{H}|^2 |\delta_V|^2 |\oper{E}^+(\brm{r}) \ket{\Phi^*_V}|^2   + \nonumber \\
  &  \frac{e^{i\phi}}{2} \sin\theta \gamma^*_{H} \gamma_{V} \delta^*_H \delta_V \bra{\Phi^*_V}\oper{E}^-(\brm{r}) \oper{E}^+(\brm{r}) \ket{\Phi^*_H}  +  \nonumber \\
   &  \frac{e^{-i\phi}}{2} \sin\theta \gamma_{H}\gamma^*_{V}\delta_H \delta^*_V \bra{\Phi^*_H}\oper{E}^-(\brm{r}) \oper{E}^+(\brm{r}) \ket{\Phi^*_V}.  
   \label{eq:Istim}
    \end{align}
    The first two terms are proportional to the intensity of the modes $\Phi_H$ and $\Phi_V$ at the transverse plane $z$ and the last two terms arise from interference between these two modes. 
    
  \subsection{Phase conjugation of the seed vector beam}
  
Let us determine conditions for which the idler field can be considered to be the phase conjugate of the VVB seed beam.  First, let us consider an intense seed beam $|\alpha_s|^2 \gg 1$, so that the SPDC component can be ignored. In this case, the output field corresponds to a pure state. As observed in section \ref{sec:III},  the polarization of the pump beam plays a crucial role in phase conjugation of the polarization state of the seed beam. Following these results, choosing the pump beam as linear-diagonally polarized so that $\gamma_H=\gamma_V$,  the idler field is described by 
\begin{equation}
\ket{\psi_i}  \approx \delta^*_H \ket{H,\Phi^*_H} + \delta^*_V \ket{V,\Phi^*_V}.  
\label{eq:psi5}
\end{equation}

The presence of the complex conjugate of the polarization coefficients $\delta^*_{H(V)}$ shows that this degree of freedom is conjugated with respect to the seed beam. We can also see that the spatial modes $\Phi^*_{H(V)}$ defined in Eq.\,\eqref{eq:stimmodes} depend on the phase conjugated amplitude of the seed . However, it still depends on the pump beam amplitude. 
Inspection of the these states indicates that they correspond to phase conjugation of the transverse spatial modes in the limiting case in which the pump beam can be approximated as a plane wave, and the crystal is very thin, so that we have ${\Psi}_{\bar{j}}(\brm{q}_s,\brm{q}_i)  = \delta (\brm{q}_s+\brm{q}_i)$:  
\begin{equation}
\ket{j,\Phi^*_j} = \int  d\brm{q}_i \phi^*_j(-\brm{q}_i) \ket{j,\brm{q}_i} \equiv \ket{j,\phi^*_j},  
\end{equation}
where the idler mode is mirror reflected ($\brm{q}_i\rightarrow -\brm{q}_i$), which can be absorbed into definition of the coordinate system. 

Then, the idler field is described by
\begin{equation}
\ket{\psi_i}  \approx \delta^*_H \ket{H,\phi^*_H} + \delta^*_V \ket{V,\phi^*_V}.  
\label{eq:psi6}
\end{equation}

This is the phase conjugate of a single photon vector beam of Eq.\,\eqref{eq:1vbpc}, showing that the idler photon of Eq.\,\eqref{eq:psi6} is in a vector vortex mode that is exactly given by the phase conjugate of the seed VVB mode.  The condition ${\Psi}_{\bar{j}}(\brm{q}_s,\brm{q}_i)  = \delta (\brm{q}_s+\brm{q}_i)$ corresponds to a SPDC source that produces maximal spatial entanglement \cite{fedorov07,walborn07c,straupe11}, so it is apparent that high-fidelity phase conjugation of VVB is obtained when the source is capable of producing both high-quality polarization and spatial entanglement.    
\par
Let us further explore the phase conjugated output.  The thin-crystal approximation is valid when the crystal length $d \ll z_R$ \cite{walborn05b}, where $z_R$ is the Rayleigh range of the pump laser.  Under these conditions, the two-photon amplitudes are determined by the transverse spatial modes of the pump beam: ${\Psi}_{\bar{j}}(\brm{q}_s,\brm{q}_i)  = \psi_j (\brm{q}_s+\brm{q}_i)$. Thus, the idler states become 
\begin{equation}
\ket{j,\Phi^*_j} = \int  d\brm{q}_i \tau_j(\brm{q}_i) \ket{j,\brm{q}_i},  
\end{equation}
with the mode functions given by the convolution
\begin{equation}
\tau_j(\brm{q}_i) = \int d\brm{q}_s  {\psi}_{\bar{j}}(\brm{q}_s+\brm{q}_i) \phi^*_j(\brm{q}_s). 
\end{equation}
Here we see that the conjugated spatial modes in Eq.\,\eqref{eq:psi5} can be tailored by manipulating those of the pump beam. Under these conditions, the intensity distribution of the phase conjugated output ($\gamma_H=\gamma_V$) after free propagation and projection onto the $\ket{\theta,\phi}$ polarization state can be obtained from Eq.\,(\ref{eq:Istim}), giving
\begin{align}
I_{\theta,\phi}^{stim}(\brm{r}_i) = &
\frac{1}{2}\cos^2\frac{\theta}{2}  |\delta_H|^2 |\mathcal{F}_H(\boldsymbol{\rho_i})|^2   + \nonumber  \\
 & \frac{1}{2} \sin^2\frac{\theta}{2}  |\delta_V|^2 |\mathcal{F}_V(\boldsymbol{\rho_i})|^2   + \nonumber \\
  &  \frac{e^{i\phi}}{4} \sin\theta  \delta_V\delta^*_H \mathcal{F}^*_V(\boldsymbol{\rho_i})\mathcal{F}_H(\boldsymbol{\rho_i})  +  \nonumber \\
   &  \frac{e^{-i\phi}}{4} \sin\theta  \delta_H \delta^*_V\mathcal{F}^*_H(\boldsymbol{\rho_i})\mathcal{F}_V(\boldsymbol{\rho_i}),  
   \label{eq:Istim2}
    \end{align}
   where $\boldsymbol{\rho}=(x,y)$,
   \begin{equation}
\label{vector99f}
\mathcal{F}_k(\boldsymbol{\rho_i}) =  \int d\brm{\rho} \,\, {\cal W}_{\bar{k}}(\boldsymbol{\rho}) {\cal U}^{\ast}_{k}(\boldsymbol{\rho})\times \mbox{exp} \left[ i|\boldsymbol{\rho}_i - \boldsymbol{\rho}|^2\frac{k_i}{2z}\right],
\end{equation}
and ${\cal W}_{\bar{k}}(\boldsymbol{\rho})$, ${\cal U}^{\ast}_{k}(\boldsymbol{\rho})$ are the transverse mode profiles of the pump beam and seed beam, respectively. 

\begin{figure}
    \centering
    \includegraphics[width=\columnwidth]{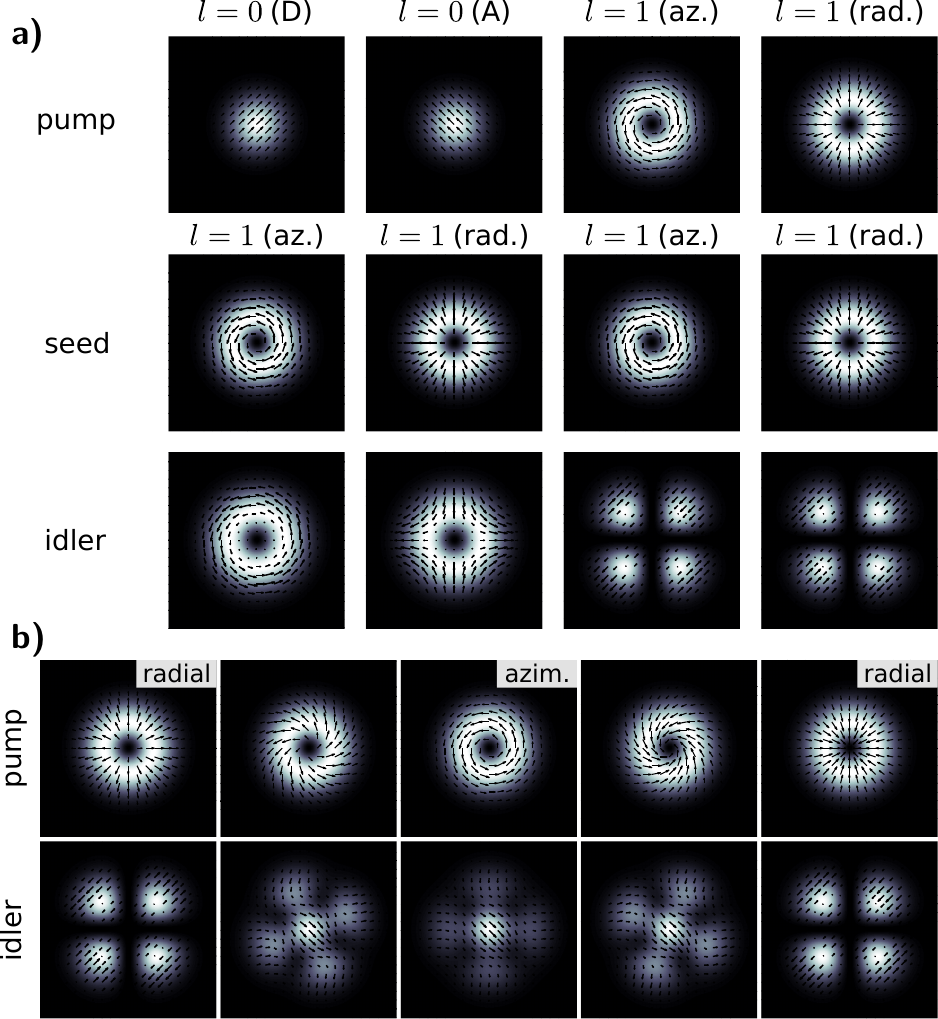}
     \caption{Simulations of idler beams for a few combinations of pump and seed. (a) Transverse profiles of pump (top panel) and seed (middle panel) beams, and corresponding simulated idler beam profile (bottom panel). The seed is a vector vortex beam, while the pump takes linear and vector polarizations. (b) For a radial seed beam, the pump beam varies from radial to azimuthal with intermediate configurations. Both pump and calculated idler profiles are shown. For all subfigures, the colorscale indicates the intensity, and the arrows, the polarization.} 
     \label{fig:intensity}
\end{figure}

In Fig.\,\ref{fig:intensity} we present use Eqs.\,\eqref{eq:Istim2} and \eqref{vector99f} to produce numerical simulations for the intensity profiles and vector structure of simulated idler beams, for a few combinations of pump and seed beams.  The dependence of the idler beam on both pump and seed parameters is clearly present.

\section{High-order Stokes parameters and high-order Poincar\'e Spheres}
\label{sec:V}
In previous works, we have drawn a geometrical picture of phase conjugation for scalar orbital angular momentum (OAM) modes \cite{Oliveira19} and for polarization states  \cite{Oliveira20}, using the Poincar\'e sphere. Let us now turn to a convenient formalism for visualizing  phase conjugation of a special case of VVBs using higher-order Stokes parameters and corresponding Poincar\'e spheres.  
\subsection{Cylindrical Vector modes as two-dimensional systems}

Cylindrically symmetric vector (CSV) modes (propagation in the $z$-direction) have the form
\begin{equation}\label{eq:CVmodes}
    \ket{\Psi_{CSV}} = \gamma_1\ket{R,+l}+\gamma_2\ket{L,-l},
\end{equation}
where $\ket{...\,,l}$ represents a state with spatial mode carrying a winding number $l=0,\pm 1,\pm 2,\ldots$ associated with OAM. The complex coefficients  $\gamma_{1,2}$ are  subject to the normalization condition ($|\gamma_1|^2+|\gamma_2|^2=1$). In most cases the spatial component of the CSV modes is described by Laguerre-Gaussian transverse modes ($LG_{lp}$) with null radial number $p$:
\begin{equation}
    LG_{l0}(\rho,\phi,z)=U_{|l|}(\rho,z)\,e^{-i l\phi}.
    \label{eqLag}
\end{equation}

\begin{figure}
    \centering
    \includegraphics[width=\columnwidth]{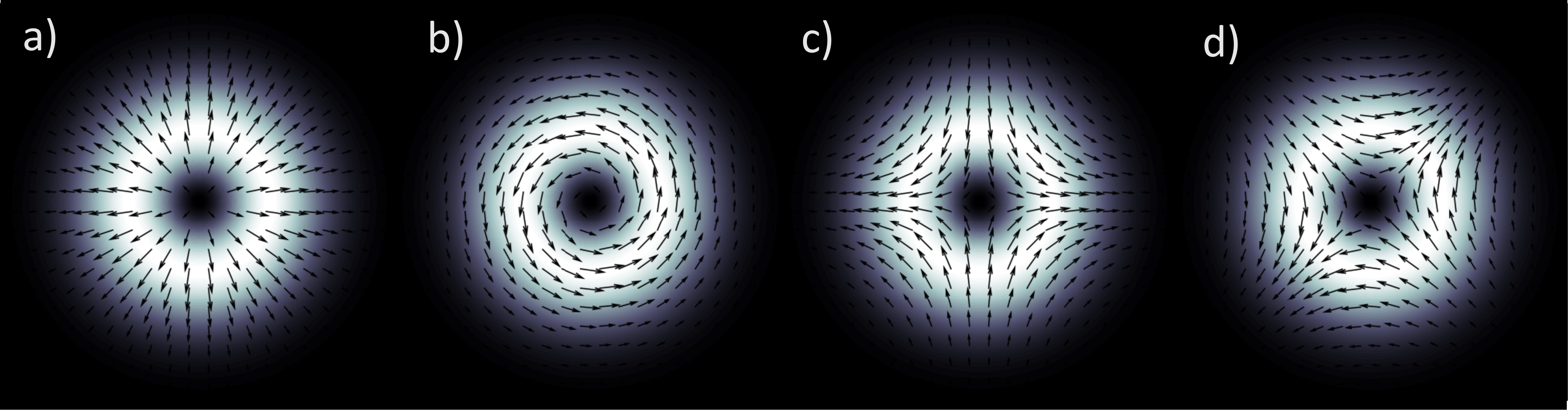} 
     \caption{Intensity and polarization profiles for $\gamma_1=\gamma_2$ and $\gamma_1=-\gamma_2$, respectively, for a) and b) $l=+1$ and c) and d) $l=-1$.} 
       \label{fig:vectorbeams}
\end{figure}

Since $U_{|l|}(\rho,z)$ does not depend on the sign of the OAM, in a simplified picture the kets in Eq.\,(\ref{eq:CVmodes}) represent the azimuthal phase and polarization modes:
\begin{eqnarray}
\label{eq:cvbasis}
\ket{R,+l} \sim e^{-il\phi}(\mathbf{e}_x+i\mathbf{e}_y); \\ \nonumber
\ket{L,-l} \sim e^{+il\phi}(\mathbf{e}_x-i\mathbf{e}_y).
\end{eqnarray}
For $l\neq 0$, Eq.\,(\ref{eq:CVmodes}) defines two sets of states for each order $|l|$ depending on the sign of $l$. For example, for $|l|=1$, the set for $l=+1$ consists of rotationally symmetric vector modes such as the radial ($\gamma_1=\gamma_2$) and azimuthal ($\gamma_1=-\gamma_2$) modes of Figs.\,\,\ref{fig:vectorbeams} a) and \ref{fig:vectorbeams} b). This set of modes are especially important for tighter beam focusing \cite{zhan09} and alignment-free optical communications \cite{dambrosio12}. Figures\,\ref{fig:vectorbeams} c) and \ref{fig:vectorbeams} d) represent two anti-vortex modes ($\gamma_1=\gamma_2$ and $\gamma_1=-\gamma_2$, respectively) contained in the $l=-1$ set. 

One may notice that, for a given $l$, the CSV modes contained in the set can be spanned by a two-dimensional basis
\begin{equation}\label{eq:compbasis}
\{\ket{0_l}=\ket{R,l},\ket{1_l}=\ket{L,-l}\},
\end{equation}
so that, ignoring the spatial amplitude $U_{|l|}$,
\begin{equation}\label{eq:cvcomp}
\ket{\psi_{CSV}}=\cos{(\theta/2)}\ket{0_l}+e^{i\varphi}\sin{(\theta/2)}\ket{1_l}    
\end{equation}
represents a general CSV mode of the set. It is important to notice that we need two distinct two-dimensional spaces to describe all the CSV modes of a given order $|l|$, depending on the sign of $l$. Of course, other bases can be constructed by linear combinations of the basis states of Eq.\,(\ref{eq:compbasis}).

\subsection{HOSPs and HOPS}

In the formalism of density matrices, two-dimensional CSV states of a given $l$ can be spanned in terms of Pauli matrices ($\hat{\sigma}_1\equiv\hat{\sigma}_z$, $\hat{\sigma}_2\equiv\hat{\sigma}_x$ and $\hat{\sigma}_3\equiv\hat{\sigma}_y$):
\begin{equation}\label{eq:cvdens}
    \hat{\rho}^l=\frac{1}{2}\sum_{n=0}^3 S_n^l\hat{\sigma}_n,
\end{equation}
weighted by the so-called high-order Stokes parameters (HOSPs)\,\,\cite{milione2011higher,milione2012higher} 
\begin{equation}
S_n^l=\text{tr}[\hat{\rho}^l\hat{\sigma}_n],    
\end{equation}
which reduce to the usual Stokes parameters for $l=0$. In terms of basis states of Eq.\,\eqref{eq:compbasis}, these parameters are
\begin{align}
S_0^l&=|\braket{0_l}{\psi_{CSV}}|^2+|\braket{1_l}{\psi_{CSV}}|^2,  \\ 
S_1^l&=2\,\text{Re}\left(\braket{0_l}{\psi_{CSV}}^*\braket{1_l}{\psi_{CSV}}\right), \nonumber \\ 
S_2^l&=2\,\text{Im}\left(\braket{0_l}{\psi_{CSV}}^*\braket{1_l}{\psi_{CSV}}\right), \nonumber\\
S_3^l&=|\braket{0_l}{\psi_{CSV}}|^2-|\braket{1_l}{\psi_{CSV}}|^2 \nonumber, 
\end{align}
and, as usual, $S_0^l=\sqrt{(S_1^l)^2+(S_2^l)^2+(S_3^l)^2}\le1$, with the equality holding only for completely polarized light.

\begin{figure}
    \centering
    \includegraphics[width=\columnwidth]{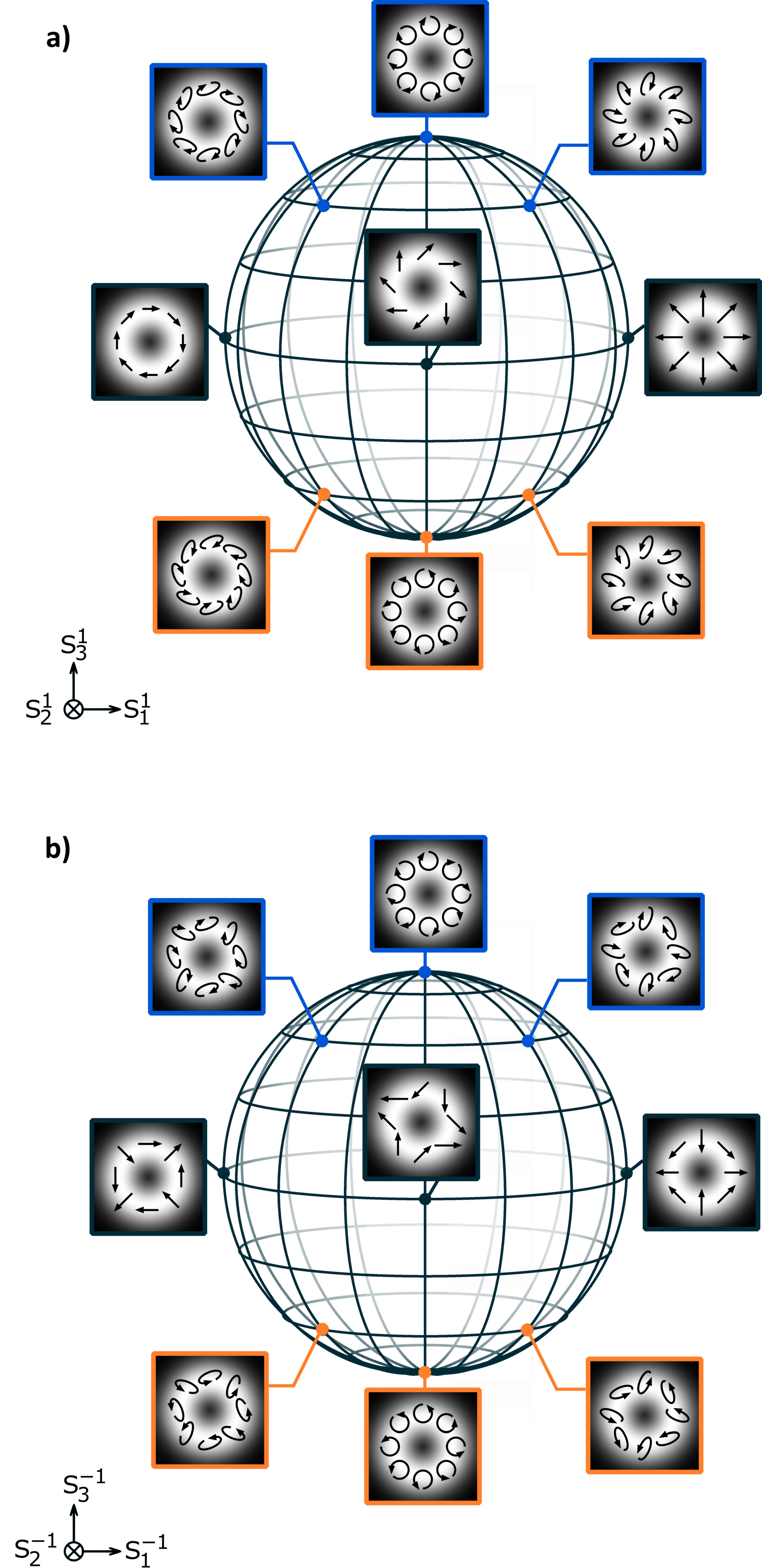} 
     \caption{(Color online) High-order Poincar\'e spheres for a) $l=+1$ ; b) $l=-1$.}
       \label{fig:HOPS}
\end{figure}

The mapping of CSV modes in two-dimensional spaces admits a geometrical representation in terms of high-order Poincar\'e spheres (HOPS) \cite{milione2011higher,milione2012higher,suzuki15}, constructed by using $S_1^l$, $S_2^l$ and $S_3^l$ as cartesian coordinates, or, equivalently, by spherical coordinates defined by
\begin{align}
r & = S_0^l, \nonumber \\ 
\theta & = \cos^{-1}(S_3^l/S_0^l) \nonumber \\ 
\varphi & = \tan^{-1}(S_2^l/S_1^l).
\end{align}
Notice that the HOPS representation is simply the Poincar\'e sphere representation for states of Eqs.\,(\ref{eq:cvcomp}) and (\ref{eq:cvdens}). Of course, two spheres are needed to represent the two sets of modes of a given order $|l|$. In Fig.\,\ref{fig:HOPS} the representation for usual states with $|l|=1$ is shown.  

\subsection{Phase conjugation on the HOPS}
Let us now investigate the phase conjugation of CSV modes and how we can interpret it geometrically on the HOPS. Since the operation performed by our two-crystal StimPDC source depends on the properties of the pump beam, let us consider the more usual case where the phase conjugation implies in changing $l\rightarrow-l$ \cite{caetano02,Oliveira19} and $R\leftrightarrow L$ \cite{Oliveira20}. One may see by expressions in Eqs.\,(\ref{eq:cvbasis}) and (\ref{eq:cvcomp}) that        
\begin{eqnarray}
\ket{0_l}^*=e^{+il\phi}(\mathbf{e}_x-i\mathbf{e}_y)=\ket{1_l}; \\ \nonumber
\ket{1_l}^*=e^{-il\phi}(\mathbf{e}_x+i\mathbf{e}_y)=\ket{0_l}.
\label{conj}
\end{eqnarray}

Equations\,(\ref{conj}) show that the conjugate counterpart of a general CSV beam of Eq.\,(\ref{eq:cvcomp}) is given by
\begin{equation}
\ket{\psi_{CSV}}^*=\cos{(\theta/2)}\ket{1_l}+e^{-i\varphi}\sin{(\theta/2)}\ket{0_l},    
\end{equation}
which in turn can be rewritten as:
\begin{equation}\label{eq:cvconjugate}
\ket{\psi_{CSV}}^*=\cos{\left(\frac{\pi-\theta}{2}\right)}\ket{0_l}+e^{i\varphi}\sin{\left(\frac{\pi-\theta}{2}\right)}\ket{1_l}.
\end{equation}
The expression in Eq.\,(\ref{eq:cvconjugate}) shows that conjugate beams are connected by reflection across the equatorial plane of the HOPS (which can be seen as changing $S_3^l\rightarrow-S_3^l$), which agrees with the $l=0$ case. Thus, we see that conjugate counterparts belong to the same HOPS as the original CSV, illustrating that vector beam conjugation does not alter the nature of the correlations between spatial and polarization DoFs, since changing spheres would imply changing correlation to anti-correlation and vice-versa. Moreover, we see that radially polarized  beams \cite{zhan09} or rotationally-invariant ($l=1$) \cite{dambrosio12} beams, do not lose their interesting properties under phase conjugation. 

\section{Phase conjugation of vector beams: generalized Stokes parameters}
\label{sec:VI}

While describing CSV modes in two-dimensional spaces gives us a practical geometrical tool for representing the modes and interpreting vector phase conjugation, we cannot represent all possible CSV states for a given order in this way, since we cannot represent a generic statistical mixture or even a superposition of modes in different two-dimensional subspaces as defined the last two sections. Moreover, one can also create a vector beam that is a superposition of modes with different values of $|l|$.  For a general description of phase conjugation of VVBs in terms of Stokes parameters, we employ a description inspired by the generalized Bloch vector for two qubits. First, since the polarization DoF is two-dimensional, Schmidt decomposition of a vector beam can always be represented by a $2 \times 2$ system, in analogy with the case of two qubits.  

For a two-qubit system, the density operator can be spanned as \cite{jaeger07}
\begin{equation}
\hat{\rho}= \frac{1}{4}\sum_{i,j=0}^3 S_{ij}\, (\hat{\sigma}_i \otimes \hat{\sigma}_j),
\end{equation}
which is equivalent to expression in Eq.\,(\ref{eq:cvdens}) generalized for a 4-dimensional system.  

The first qubit corresponds to the polarization DoF and the second to the transverse spatial mode. In this way, we can write the Stokes matrix as  
\begin{equation}
\mathbb{S}  =
\left ( 
\begin{matrix}
1 & s_{01} & s_{02}  & s_{03}  \\ 
s_{10} & s_{11} & s_{12}  & s_{13}  \\ 
s_{20} & s_{21} & s_{22}  & s_{23}  \\ 
s_{30} & s_{31} & s_{32}  & s_{33}  \\ 
\end{matrix}
\right )
\end{equation}
where we used $s_{00}=1$, corresponding to normalized components. Note that $s_{0j}$ and $s_{j0}$ ($j=1,2,3$) are the Stokes vector components of the individual qubits, while the $s_{jk}$ components contain the correlations.

\begin{figure}
\vspace{0.5cm}
    \centering
    \includegraphics[width=\columnwidth]{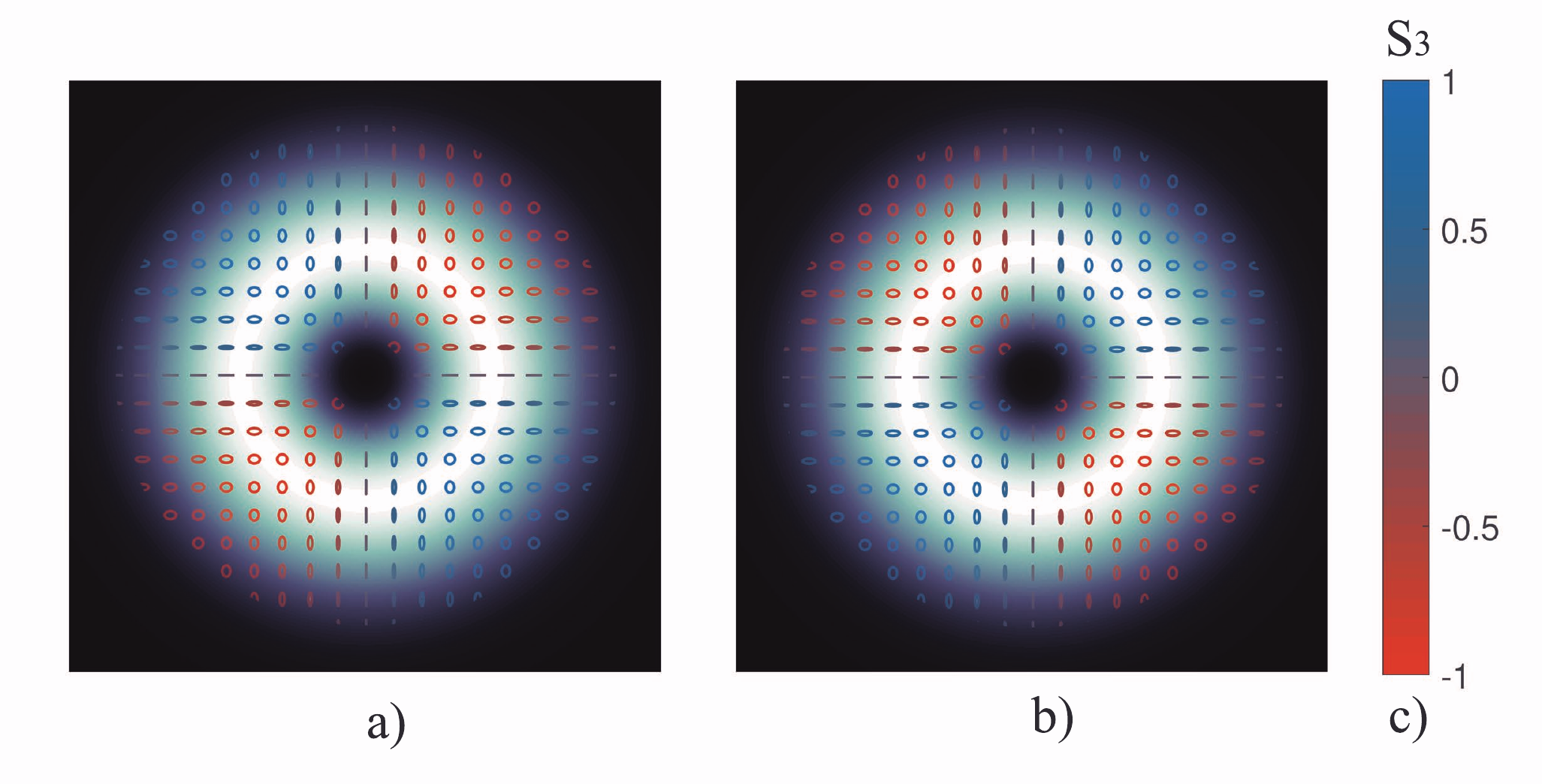}
     \caption{a) Anisotropic CSV beam and b) phase conjugate beam. c) Color bar for the $S_3$ Stokes parameter for polarization.}
       \label{fig:cvaniso}
\end{figure}

Using the representation $\hat{\sigma}_1\equiv\hat{\sigma}_z$, $\hat{\sigma}_2\equiv\hat{\sigma}_x$ and $\hat{\sigma}_3\equiv\hat{\sigma}_y$ as in the last section, phase conjugation of the vector beam changes the sign of the third component of both qubits, thus 
\begin{equation}
\mathbb{S}^*  =
\left ( 
\begin{matrix}
1 & s_{01} & s_{02}  & -s_{03}  \\ 
s_{10} & s_{11} & s_{12}  & -s_{13}  \\ 
s_{20} & s_{21} & s_{22}  & -s_{23}  \\ 
-s_{30} & -s_{31} & -s_{32}  & s_{33}  \\ 
\end{matrix}
\right ).
\end{equation}
For example, in Ref. \cite{Oliveira20} an anisotropic vector beam was produced, which can be represented by
\begin{equation}
   \ket{\psi}_{\text{aniso}}=\ket{D,+|l|}+\ket{A,-|l|}
\end{equation}
where $\ket{D(A)} = (\ket{H} +(-) \ket{V})/\sqrt{2}$ are diagonal (antidiagonal) linear polarization states. In Fig.\,\ref{fig:cvaniso}a) one can see a plot of this beam, which
can also be represented as
\begin{equation}
\mathbb{S}_\text{aniso}  =
\left ( 
\begin{matrix}
1 & 0 & 0  &0  \\ 
0 & 1 & 0  & 0  \\ 
0 & 0 & 0  & 1  \\ 
0 & 0 & 1  & 0  \\ 
\end{matrix}
\right )
\end{equation}
 and the Stokes matrix of the conjugate beam (Fig.\,\ref{fig:cvaniso}b)) is 
 \begin{equation}
 \mathbb{S}^*_\text{aniso}  =
\left ( 
\begin{matrix}
1 & 0 & 0  &0  \\ 
0 & 1 & 0  & 0  \\ 
0 & 0 & 0  & -1  \\ 
0 & 0 & -1  & 0  \\ 
\end{matrix}
\right ),
\end{equation}
which agrees with the experimental results in Ref.\,\cite{Oliveira20}.

\section{Conclusion}
\label{sec:VII}
In conclusion, we have introduced a theoretical approach for the description of StimPDC including the use of vector vortex beams, which is based on a quantum-optics formalism. In addition to general expressions showing the interplay between the polarization and spatial DoFs of the pump and seed (signal) beam, we describe the conditions for which the stimulated idler field is the phase conjugate of the input seed beam.  We also make use of higher-order Stokes parameters and Poincar\'e sphere (HOSPs and HOPS) to provide a geometrical interpretation for phase conjugation of cylindrically symmetric vortex beams in StimPDC, and provide a general Stokes matrix description of phase conjugation that is applicable to any VVB. 
\par
The results presented here are a contribution to the field of nonlinear optics with vectorial and structured light. We note that recent work has studied polarization control and transverse mode mixing in up-conversion with single type-II crystals \cite{pereira17,buono18}, as well as up-conversion of VVBs using a type-II crystal entanglement source \cite{wu2020polarization}.
We believe that both the combination of StimPDC and VVB and our theoretical approach are very useful for a broad range of applications and experimental investigations.
A possible application of our results is more efficient quantum state tomography of SPDC sources of entangled vortex modes or hyper-entangled states involving polarization and transverse modes \cite{barreiro10}, or in determination of symmetry properties of nonlinear materials \cite{rocio16} .  The nonclassical properties of the multi-mode output of the signal beam could also be explored. Moreover, phase conjugation of VVBs could be used to mediate detrimental propagation effects due to atmospheric turbulence and other causes.  An experiment exploring this capabilities of StimPDC and VVB  is currently in progress in our Laboratory. 
\begin{acknowledgements}
The authors thank the Brazilian Agencies CNPq, FAPESC, FAPERJ (E- 26/010.002997/2014 and E-26/202.7890/2017), and the Brazilian National Institute of Science and Technology of Quantum Information (INCT/IQ). This study was funded in part by the Coordena\c{c}\~{a}o de Aperfei\c{c}oamento de Pessoal de N\'{i}vel Superior - Brasil (CAPES) - Finance Code 001. SPW received support from the Chilean Fondo Nacional de Desarrollo Científico y Tecnológico (Conicyt) (1200266) and Millennium Institute for Research in Optics. 
\end{acknowledgements}

\bibliographystyle{apsrev}

\end{document}